\documentclass[twocolumn,preprintnumbers,superscriptaddress,aps,nofootinbib,prl,10pt]{revtex4-2}
\usepackage{psfrag,slashed,cancel,array,graphicx,todonotes,hyperref}
\usepackage{subfigure}
\usepackage[labelfont=bf,labelsep=quad,justification=justified]{caption}
\usepackage[utf8]{inputenc}
\usepackage{mathtools}
\usepackage{mathrsfs}
\usepackage{multirow}
\usepackage{braket}
\usepackage{slashed}
\usepackage{booktabs} 
\usepackage[normalem]{ulem}
\usepackage{slashed}
\hypersetup{pdftitle={},pdfcreator={},linkcolor=[rgb]{0.15,0.35,0.75},colorlinks=true,citecolor=[rgb]{0.675,0,0.2},urlcolor=[rgb]{0.15,0.35,0.65}}

\usepackage{ragged2e}
\makeatletter
\let\original@makecaption\@makecaption
\def\@makecaption#1#2{%
  \original@makecaption{#1}{\justifying #2\par}%
}
\makeatother
\def\beq{\begin{equation}}
\def\eeq{\end{equation}}
\def\bsp#1\esp{\begin{split}#1\end{split}}

\newcommand{\lqcd}{\Lambda_\mathrm{QCD}}

\newcommand{\as}{\alpha_s}

\AtBeginDocument{
\heavyrulewidth=.08em
\lightrulewidth=.05em
\cmidrulewidth=.03em
\belowrulesep=.65ex
\belowbottomsep=0pt
\aboverulesep=.4ex
\abovetopsep=0pt
\cmidrulesep=\doublerulesep
\cmidrulekern=.5em
\defaultaddspace=.5em
}

\def\be{\begin{equation}}
\def\ee{\end{equation}}

\begin{document}

\preprint{CERN-TH-2026-173}

\title{Third-Order Logarithmic Resummation for Jet-Vetoed Higgs Production}

\author{Samuel Abreu}
\email{samuel.abreu@ugr.es}
\affiliation{CERN, Theoretical Physics Department, CH-1211 Geneva 23, Switzerland}
\affiliation{Higgs Centre for Theoretical Physics, School of Physics and Astronomy,
 The University of Edinburgh, Edinburgh EH9 3FD, Scotland, United Kingdom}
\affiliation{
Departamento de F\'isica Te\'orica y del Cosmos, Universidad de Granada, Campus de Fuentenueva, E-18071 Granada, Spain}
\author{Jonathan R.~Gaunt}
\email{gaunt.jonathan@ucy.ac.cy}
\affiliation{Department of
  Physics and Astronomy, University of Manchester, Manchester M13 9PL,
  United Kingdom}
\affiliation{Department of Physics, University of Cyprus, Nicosia 1678, Cyprus}
\author{Pier Francesco Monni}
\email{pier.monni@cern.ch}
\affiliation{CERN, Theoretical Physics Department, CH-1211 Geneva 23, Switzerland}
\author{Luca Rottoli}
\email{luca.rottoli@unimib.it}
\affiliation{Dipartimento di Fisica G. Occhialini, Universit\`a degli Studi di Milano-Bicocca and INFN,
Sezione di Milano-Bicocca, Piazza della Scienza 3, 20126 Milano, Italy}
\author{Robert Szafron}
\email{rszafron@bnl.gov}
\affiliation{Department of Physics, Brookhaven National Laboratory,
Upton, N.Y., 11973, U.S.A.}

\begin{abstract}
Jet vetoes are central to precision measurements of Higgs-boson
properties at the Large Hadron Collider, where they are used to suppress
backgrounds containing hard QCD radiation. We present the first prediction for the Higgs jet-vetoed cross section at the third logarithmic order in QCD, consistently combining the next-to-next-to-next-to-leading order calculation with state-of-the-art next-to-next-to-next-to-leading logarithmic resummation. At phenomenologically relevant veto scales ($p_T^{\rm v}\sim30$ GeV) and jet radii ($R=0.4$), the resummation increases the fixed-order prediction by about $3\%$, and their combination leads to a residual perturbative uncertainty of approximately $3\%$.
Our results mark a milestone in the precise description of jet-vetoed Higgs-production at hadron colliders, reaching the level of precision relevant for the high-luminosity phase of the LHC.
\end{abstract}

\maketitle

\paragraph*{Introduction.---}
Jet-vetoed cross sections are central to the measurement of Higgs-boson couplings at the LHC. For instance, they are employed to suppress QCD background originating from top-quark decays in $H\to WW^*$~\cite{CMS:2022uhn,CMS:2026igg,ATLAS:2025hki} and $H\to\tau\tau$~\cite{CMS:2021gxc,CMS:2022kdi,ATLAS:2024wfv} analyses, as well as in the definition of Simplified Template Cross Section jet bins~\cite{ATLAS:2022ooq,CMS:2022uhn,LHCHiggsCrossSectionWorkingGroup:2016ypw}.
The zero-jet bin is defined by demanding the absence of any jets with a transverse momentum above a veto scale $p_T^{\rm v}$ and provides an exceptionally clean environment for precision measurements, with projections for the corresponding measurements at the high-luminosity LHC (HL-LHC) reaching the $3$--$5\%$ level~\cite{CMS:2025HLHiggs,ATLAS:2025HLHiggs}.
Matching this level of precision in theoretical predictions is a non-trivial task. In particular, the introduction of a veto scale severely restricts the phase space available for real QCD radiation, which induces large Sudakov logarithms $\ln(M/p_T^{\rm v})$ at each order in perturbation theory ($M$ denotes the Higgs-boson mass). In the kinematic region relevant for the LHC, these logarithms become large enough to spoil the convergence of the standard fixed-order expansion, necessitating an all-orders resummation.
The resummation of jet-veto logarithms has a long history~\cite{Banfi:2012yh,Banfi:2012jm,Becher:2012qa,Boughezal:2013oha,Becher:2013xia,Stewart:2013faa,Banfi:2013eda,Becher:2014aya,Monni:2019yyr,Gavardi:2023aco,Campbell:2023cha}. The next-to-next-to-leading logarithmic (NNLL) accuracy was originally achieved in Refs.~\cite{Banfi:2012jm,Becher:2013xia,Stewart:2013faa}, and later combined with fixed-order predictions through next-to-next-to-next-to-leading order (N$^3$LO)~\cite{Banfi:2015pju}. 

In this Letter, we present the first prediction for the jet-vetoed Higgs production cross section at next-to-next-to-next-to-leading accuracy in both logarithmic and fixed-order perturbation theory, namely N$^3$LL$+$N$^3$LO. 
The N$^3$LL calculation implies a resummation of terms ${\cal O}(\alpha_s^n\ln^{m\geq n-2}(M/p_T^{\rm v}))$ ($\alpha_s$ is the strong coupling constant) in the logarithm of the cross section, and builds upon a set of recently determined perturbative ingredients~\cite{Abreu:2022sdc,Abreu:2022zgo,companion} in the framework of soft-collinear effective field theory (SCET)~\cite{Bauer:2000yr,Bauer:2001yt,Bauer:2002nz,Beneke:2002ph,Beneke:2002ni}. 
As we discuss below, we find that the N$^3$LL corrections are sizable. For instance, they increase the cross section by $\sim\!3\%$ compared to the N$^3$LO result at $p_T^{\rm v}=30$~GeV, highlighting their importance for precision Higgs phenomenology.
The size of the corrections is consistent with the observation of Ref.~\cite{companion} that the three-loop rapidity anomalous dimension is large.

The results obtained in this Letter represent a milestone in the Higgs precision program, reaching the level of accuracy foreseen for HL-LHC for the jet-veto cross section~\cite{CMS:2025HLHiggs,ATLAS:2025HLHiggs}. Furthermore, they enable analogous resummations in a range of other key processes, from the production of more complex color-singlet systems to heavy-quark final states or jet processes~\cite{Liu:2012sz,Cal:2024yjz,vanBeekveld:2025zjh}.
Finally, jet-veto resummation finds application also in the context of infrared slicing for higher-order calculations~\cite{Abreu:2022zgo} and the matching of NNLO calculations to parton-shower simulations~\cite{Gavardi:2023aco}, opening the way to a broader class of precision QCD predictions at colliders.

\paragraph*{Factorization of the jet-vetoed cross section.---}
In the limit $p_T^{\rm v}\ll M$, the differential cross section for Higgs production in gluon fusion at hadron colliders with a veto on the leading-jet transverse momentum admits a factorization formula in SCET that reads~\cite{Becher:2012yn,Becher:2013xia,Stewart:2013faa}:
\begin{align}\label{eq:SCET-fact}
    \frac{d\sigma(p_T^{\rm v})}{dy} &=  \hat{\sigma}_{\rm B}
    \, \mathcal{H}(M;\mu)\,
      \mathcal B_{n,g}(x_{n}, p_T^{\rm v}, R^2;\mu,\nu)\notag\\
    &\!\!\!\times \mathcal B_{\bar{n},g}(x_{\bar{n}}, p_T^{\rm v}, R^2;\mu,\nu)\,
      \mathcal S(p_T^{\rm v}, R^2;\mu,\nu)\,.
\end{align}
Here,
$\hat{\sigma}_{\rm B}=\frac{M^2\,\alpha_s^2(\mu)}{72\pi\left(N_c^2-1\right)s\,v^2}$ is the Born-level partonic cross-section for the production of a Higgs boson in the heavy-top effective theory, which depends on the hadronic center of mass energy $s$ and the Higgs field vacuum expectation value $v=246.2$ GeV, $x_{n,\bar{n}} = (M/\sqrt{s})\,e^{\pm y}$ with $y$ the rapidity of the Higgs boson, $R$ is the jet radius, and $\mu$, $\nu$ are the renormalization and rapidity regularization scales; for the latter, we use the scheme of Ref.~\cite{Li:2016axz}. $N_c$  is the number of colors in QCD, $N_c=3$.
The jet-veto constraint on the jet with the largest transverse momentum is imposed using jet algorithms from the generalized-$k_t$ family~\cite{Cacciari:2008gp,Dokshitzer:1997in,Catani:1993hr}. 
The factorization~\eqref{eq:SCET-fact} may be violated starting at five loops due to coherence violating logarithms~\cite{Forshaw:2006fk,Forshaw:2008cq,Forshaw:2009fz,Banfi:2010xy,DuranDelgado:2011tp,Gaunt:2014ska,Zeng:2015iba,Forshaw:2021fxs,Becher:2021zkk,Becher:2023mtx,Boer:2024hzh,Banfi:2025mra,Dasgupta:2025cgl,Becher:2026kbr}, which we do not consider here.

The \emph{hard function} $\mathcal{H}(M;\mu) = |{ C}_S(M;\mu)|^2$  encodes the virtual 
contributions at the hard scale $\mu \sim M$, and is known through three loops in QCD~\cite{Moch:2005tm,Baikov:2009bg,Gehrmann:2010ue}.
The two gluon \emph{beam functions} $\mathcal B_{n/\bar{n},g}$ describe collinear initial-state radiation 
along the two beam directions and are matched, in the limit $p_T^{\rm v} \gg \Lambda_{\rm QCD}$, onto collinear parton distribution functions $f_i(x,\mu)$ with matching kernels ${\cal I}_{ij}$
\begin{align}
&    \mathcal B_{n/\bar{n},i}(x_{n/\bar{n}}, p_T^{\rm v}, R^2;\mu,\nu)= \left(\mathcal I\otimes f\right)_i\\&\!\equiv\int_{x_{n,\bar{n}}}^1 \!\!\!\frac{dz}{z} \mathcal I_{ij} (z, p_T^{\rm v}, R^2;\mu,\nu) f_j\left(\frac{x_{n/\bar{n}}}{z},\mu\right)+{\cal O}\left(\frac{\lqcd}{p_T^{\rm v}}\right),\notag
\end{align}
with $\lqcd$ being of the order of the QCD confinement scale.
These kernels have been computed to two loops for both quark- and gluon-initiated processes in Refs.~\cite{Bell:2022nrj,Abreu:2022zgo,Bell:2024epn}.
The \emph{soft function} $\mathcal S$ encodes the contribution of soft wide-angle radiation emitted from the initial-state legs, subject to the jet-veto constraint.
Its two-loop computation was carried out in Refs.~\cite{Bell:2020yzz,Abreu:2022sdc}.

The resummation is achieved by evolving each ingredient from its canonical scale to a common scale via renormalization group equations (RGEs) in $\mu$.
In addition, the beam and soft functions satisfy rapidity RGEs in $\nu$
\begin{align}
\frac{d\ln \mathcal B}{d\ln\nu} = 2\!\int_{p_T^{\rm v}}^{\mu}\!\!\frac{d\mu'}{\mu'}
    \,\Gamma_{K}(\alpha_s(\mu'))\! - \!\frac{1}{2}\gamma_\nu(\alpha_s(p_T^{\rm v}), R^2)\,,
    \label{eq:rapRGE-B}\\
    \frac{d\ln \mathcal S}{d\ln\nu} = -4\!\int_{p_T^{\rm v}}^{\mu}\!\!\frac{d\mu'}{\mu'}
    \,\Gamma_{K}(\alpha_s(\mu')) + \gamma_\nu(\alpha_s(p_T^{\rm v}), R^2)\,,
    \label{eq:rapRGE-S}
\end{align}
where we have suppressed the beam-function indices. The quantity $\Gamma_{K}$ is the cusp anomalous dimension, while the rapidity anomalous dimension $\gamma_\nu(\alpha_s, R^2)$
depends on $p_T^{\rm v}$ solely through the running coupling, while its $R$ dependence encodes the dependence on the jet radius in the jet-clustering algorithm.
Evolving the hard, beam, and soft functions from their canonical scales to the common $\mu,\nu$ scales, we obtain: 
\begin{align}\label{eq:SCET-fact-2} \nonumber
    \frac{d\sigma(p_T^{\rm v})}{dy} &=  \hat{\sigma}_{\rm B}
    \, \mathcal{H}(M;M)\,
      \mathcal B_{n,g}(x_n, p_T^{\rm v}, R^2;p_T^{\rm v},M)\\ \notag
    &\!\!\!\!\!\!\!\!\!\!\!\!\!\times \mathcal B_{\bar{n},g}(x_{\bar{n}}, p_T^{\rm v}, R^2;p_T^{\rm v},M)\,
      \mathcal S(p_T^{\rm v}, R^2;p_T^{\rm v},p_T^{\rm v}) \\ \nonumber
      &\!\!\!\!\!\!\!\!\!\!\!\!\!\times \exp\left\{ \int^{M^2}_{(p_T^{\rm v})^2}\frac{d\mu^2}{\mu^2} \left(\Gamma_K(\alpha_s(\mu)) \ln\dfrac{\mu^2}{M^2} \!- \! \gamma_V(\alpha_s(\mu))\right)\right\} \\ 
      &\!\!\!\!\!\!\!\!\!\!\!\!\!\times \exp \left\{ \ln\left(\frac{M}{p_T^{\text{v}}}\right) \gamma_\nu\left(\alpha_s(p_T^{\text{v}}), R^2\right) \right\}\,,
\end{align}
where $\gamma_V$ is the non-cusp anomalous dimension of the squared gluon form factor.

Achieving N$^3$LL accuracy requires the cusp anomalous dimension $\Gamma_{K}$ at four loops~\cite{Henn:2019swt},
the non-cusp anomalous dimension $\gamma_V$ at three 
loops~\cite{Moch:2004pa,Vogt:2004mw,Moch:2005tm,Gehrmann:2010ue,Li:2014afw}
(we use the expressions in Appendix A of Ref.~\cite{Becher:2009qa}),
the boundary conditions for $\mathcal H$, $\mathcal B_{n/\bar{n}}$, and $\mathcal S$ at two loops \cite{Moch:2005tm,Baikov:2009bg,Gehrmann:2010ue,Bell:2020yzz,Bell:2022nrj,Abreu:2022sdc,Abreu:2022zgo,Bell:2024epn},
and the rapidity anomalous dimension $\gamma_\nu^{(3)}(R^2)$ at three loops.
This final ingredient, which for the first time encodes a dependence on the specific jet algorithm, was unknown until recently.
In Ref.~\cite{companion}, it is computed from the three-loop 
jet-veto soft function and presented as a function of $R$ for all jet algorithms in the generalized-$k_t$ family.
In particular, it can be decomposed as~\cite{companion}
\begin{align}\label{eq:RAD}
    \gamma_{\nu}(\alpha_s,R^2) = \gamma_{\nu,\,{\rm RSV}}(\alpha_s) + \Delta\gamma_{\nu}(\alpha_s,R^2)\,.
\end{align}
The first term $\gamma_{\nu,\,{\rm RSV}}$ is the anomalous dimension for an auxiliary reference observable~\cite{companion}
\begin{align}\label{eq:RSV}
\gamma_{\nu,\,{\rm RSV}}(\alpha_s) = 
&\,\,\gamma_{\nu,\perp}(\alpha_s) \\
&\,\,\,\,+ \left(\frac{\alpha_s}{4\pi}\right)^3 \frac{32}{3} C_A b_0^2 \,\zeta_3 + {\cal O}(\alpha_s^4) \,, \notag \, 
\end{align}
with $b_0 \!=\! (11 C_A \!-\! 2 n_f)/3$, 
while $\gamma_{\nu, \perp} (\alpha_s)$ 
is the rapidity anomalous dimension for transverse-momentum resummation, given  at three loops in Refs.~\cite{Li:2016ctv,Vladimirov:2016dll}. The second term in~\eqref{eq:RAD} encodes the dependence on the jet algorithm
\begin{equation}~\label{eq:DeltaGamma}
    \Delta\gamma_{\nu}(\alpha_s,R^2) = \sum_{n\geq 2}\left(\frac{\alpha_s}{4 \pi}\right)^{n}\Delta\gamma^{(n)}_{\nu}(R^2)\,.
\end{equation}
To connect with the QCD resummation literature, we recast Eq.~\eqref{eq:SCET-fact-2} in the scheme of Ref.~\cite{Banfi:2012jm}
\begin{align}~\label{eq:QCD-res}
    \frac{d\sigma(p_T^{\rm v})}{d y} &= \hat{\sigma}_{\rm B} {\cal H}\left(M;M\right)\, e^{-\mathcal R\left(M,p_T^{\rm v}\right)}\,{\cal F}\left(M,p_T^{\rm v},R^2\right)\notag\\&\hspace{-0.1cm}\times\, \left({\cal C}\otimes f\right)_g(x_n,p_T^{\rm v}) \left({\cal C}\otimes f\right)_g(x_{\bar{n}},p_T^{\rm v})\,.
\end{align}
The coefficient functions ${\cal C}_{ij}$ are given in terms of the boundary condition of the soft function and the matching coefficients of the beam functions, and read
\begin{align}
    {\cal C}_{ij}&(z, \alpha_s(p_T^{\rm v}), R^2) \\&\equiv \sqrt{\mathcal S(p_T^{\rm v}, R^2;p_T^{\rm v},p_T^{\rm v})}\,\mathcal I_{ij}(z, p_T^{\rm v}, R^2;p_T^{\rm v},M)\, .\notag
\end{align}
The Sudakov radiator is defined as
\begin{equation}
\mathcal R\left(M,p_T^{\rm v}\right) \!=\!\! \int_{(p_T^{\rm v})^2}^{M^2}\!\!\!\frac{d\mu^2}{\mu^2}\left(A(\alpha_s(\mu^2))\ln\frac{M^2}{\mu^2}+B(\alpha_s(\mu^2))\right),
\end{equation}
where the coefficients $A(\alpha_s)$ and $B(\alpha_s)$ are related to the anomalous dimensions of the SCET formulation by
\begin{align}\label{eq:A-B}
    A(\alpha_s) &= \Gamma_{K} (\alpha_s) +  \frac{1}{2}\beta(\alpha_s) \frac{\partial \gamma_{\nu,\,{\rm RSV}}(\alpha_s)}{\partial\alpha_s}\,,\notag\\
    B(\alpha_s) 
    &= \gamma_V(\alpha_s) - \frac{1}{2}\gamma_{\nu,\,{\rm RSV}}(\alpha_s)\,,
\end{align}
where $\beta(\alpha_s)=d\alpha_s/d\ln \mu^2$.
 Finally, the function ${\cal F}$ encodes the dependence of $\gamma_{\nu}$ on the jet algorithm and is related to Eq.~\eqref{eq:DeltaGamma} by
\begin{align}\label{eq:FF}
   {\cal F}&\left(M,p_T^{\rm v},R^2\right) = \exp\left\{\Delta\gamma_{\nu}(\alpha_s(p_T^{\rm v}),R^2)\ln\frac{M}{p_T^{\rm v}}\right\}\,.
\end{align}

\begin{figure}
\includegraphics[width = 0.45\textwidth]{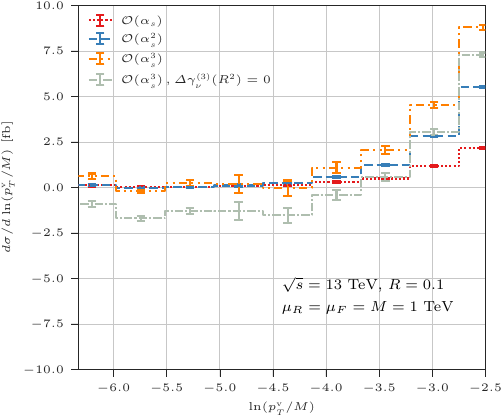}
\caption{Difference between the expansion of the derivative of Eq.~\eqref{eq:QCD-res} w.r.t.~$\ln(p_T^{\rm v}/M)$ and its fixed-order counterpart at ${\cal O}(\alpha_s)$ (red), ${\cal O}(\alpha_s^2)$ (blue) and ${\cal O}(\alpha_s^3)$ (orange). The latter is also shown with $\Delta \gamma_\nu^{(3)}(R^2)=0$ (gray).}
\label{fig:slicing}
\end{figure}

As an important test of our calculation and its implementation, we compare the difference through order $\as^k$ of the expansion of the leading-jet transverse momentum spectrum and the corresponding N$^{k-1}$LO fixed-order calculation (${\cal O}(\as^k)$ relative to the Born level) in $H + \rm jet $ production. The former is obtained by integrating Eq.~\eqref{eq:QCD-res} over $y$ and then taking the derivative with respect to~$p_T^{\rm v}$.
The difference should vanish when $p_T^{\rm v}\ll M$, since the resummed formula at N$^k$LL captures all logarithmically divergent terms up to and including ${\cal O}(\as^k)$.
In Fig.~\ref{fig:slicing} we plot the differences at order $\alpha_s$, $\alpha_s^2$ and  $\alpha_s^3$ as a function of $\ln(p_T^{\rm v}/M)$. We consider proton-proton collisions at the LHC with a center-of-mass energy of $\sqrt{s} = 13$~TeV, setting $M = 1~\rm TeV$ and $R=0.1$ as a numerical stress test since the larger hierarchy and smaller radius
enlarge the logarithms and therefore enhance sensitivity to the three-loop clustering contribution.
Eq.~\eqref{eq:QCD-res} is implemented in the \texttt{RadISH} program~\cite{Monni:2016ktx,Bizon:2017rah}, using \texttt{hoppet}~\cite{Karlberg:2025hxk} for computing convolutions and \texttt{hplog}~\cite{Gehrmann:2001pz} for efficiently evaluating harmonic polylogarithms. The fixed-order calculation for $H + \rm jet$ is obtained with the \texttt{NNLOJet} program~\cite{NNLOJET:2025rno}.\footnote{To optimize the sampling, the fixed-order results above and below $\ln(p_T^{\rm v}/M) = -5$ are obtained with separate runs.}
The difference at the three orders tends to zero, consistently with the resummation accuracy of Eq.~\eqref{eq:QCD-res}, which provides a further test of our calculation.
To quantify the impact of the rapidity anomalous dimension calculated in~\cite{companion}, we also show the same difference obtained by setting to zero the $R$-dependent contribution $\Delta \gamma_\nu^{(3)}(R^2)$ in Eq.~\eqref{eq:RAD}. 
The corresponding prediction in Fig.~\ref{fig:slicing} visibly fails to approach zero, highlighting the numerical impact of this correction.

\paragraph*{Matching to fixed order at N$^3$LL+N$^3$LO.---}
We now use Eq.~\eqref{eq:QCD-res} for the calculation of the Higgs jet-vetoed cross section at the LHC, where we integrate inclusively over the Higgs-boson rapidity $y$ and set $M$ to its physical value $M=125$ GeV.
Eq.~\eqref{eq:QCD-res} and the matching to fixed order discussed below are implemented in the \texttt{RadISH} program for the production of a generic color-singlet system. 
To estimate the perturbative uncertainties of our prediction, we restore the dependence of the resummation coefficients in Eq.~\eqref{eq:QCD-res} on the renormalization ($\mu_R$), factorization ($\mu_F$), and resummation ($\mu_L$) scales, the variation of which probes the size of higher-order corrections.
Here, we also match the resummed cross section with the N$^3$LO prediction
\begin{equation}~\label{eq:N3LO}
    \sigma_{\text{N$^k$LO}}(p_T^{\rm v}) = \sigma_{\text{N$^k$LO}}^{\rm tot.} - \sigma^{H+\rm jet}_{\text{N$^{k-1}$LO}}(p_T^j>p_T^{\rm v})\,.
\end{equation}
The total cross section, $\sigma_{\text{N$^k$LO}}^{\rm tot.}$, is known to N$^3$LO~\cite{Anastasiou:2015vya,Anastasiou:2016cez,Mistlberger:2018etf} and is computed with the \texttt{ggHiggs} program~\cite{Bonvini:2016frm} (and checked with \texttt{n3loxs} \cite{Baglio:2022wzu}), while $\sigma^{H+\rm jet}_{\text{N$^{k-1}$LO}}$ is computed through NNLO with \texttt{NNLOJet}.
For our phenomenological results, we consider proton-proton collisions at the LHC with a center-of-mass energy of $\sqrt{s} = 13.6$~TeV with $R=0.4$.
We use the NNLO PDF set \texttt{NNPDF40\_nnlo\_as\_01180}~\cite{NNPDF:2024nan} with \mbox{$\alpha_s(m_Z) = 0.118$} through the \texttt{LHAPDF} interface~\cite{Buckley:2014ana} at all perturbative orders considered in the study.
We set the pole top mass to 173.1~GeV.
To match Eqs.~\eqref{eq:QCD-res} and~\eqref{eq:N3LO}, we follow the  procedure of Ref.~\cite{vanBeekveld:2025zjh}, in which the resummation is smoothly turned off at a matching scale $p_T^{\rm v}\!\sim \!\mu_M$. We adopt the multiplicative matching formula reported in Ref.~\cite{supplemental}, through which the missing N$^3$LO leading-power non-logarithmic terms, of formal N$^4$LL order, are extracted from the fixed-order calculation.

We set the matching ($\mu_M$), renormalization ($\mu_R$), factorization ($\mu_F$) and resummation ($\mu_L$) scales to $M/2$ for our central prediction.
Missing higher-order uncertainties are estimated using a ten-point scale variation, obtained as follows: 
We vary $\mu_R$ and $\mu_F$ by a factor of two around their central values, subject to the constraint \mbox{$1/4 < \mu_R/\mu_F < 4$}. For central $\mu_R$ and $\mu_F$, we vary $\mu_L$ by a factor of $2$ and separately vary $\mu_M$ from $M/2$ to $M$. 
We also perform the matching using the alternative schemes of Refs.~\cite{Banfi:2015pju,vanBeekveld:2025zjh}, finding consistency within the uncertainties estimated as outlined above~Ref.~\cite{supplemental}.
\begin{figure}[t]
\includegraphics[width = 0.45\textwidth]{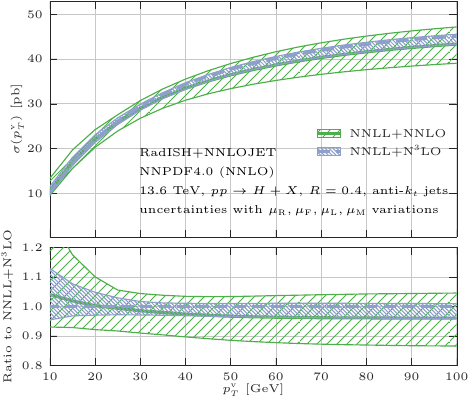}
\caption{The previous state-of-the-art for the jet-vetoed cross section: NNLL+N$^3$LO, compared to NNLL+NNLO for $\sqrt{s}=13.6~\mathrm{TeV}$ as a function of the veto scale $p_T^{\rm v}$, for anti-$k_t$ jets with $R=0.4$.}
\label{fig:N3LOcorrection}
\end{figure}

\paragraph*{The Higgs jet-vetoed cross section at the LHC.---}
%
\begin{figure*}[ht]
\includegraphics[width = 0.45\textwidth]{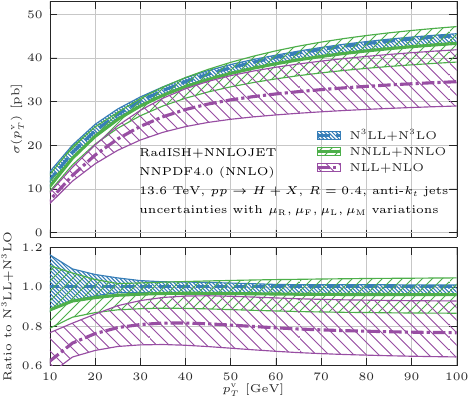}
\includegraphics[width = 0.45\textwidth]{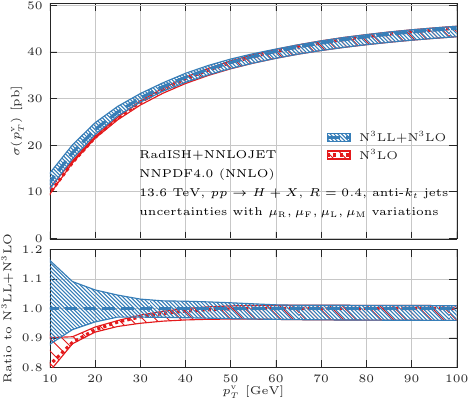}
\caption{Jet-vetoed cross section at $\sqrt{s}=13.6~\mathrm{TeV}$ as a function of the veto scale $p_T^{\rm v}$, for anti-$k_t$ jets with radius parameter $R=0.4$.
The left panel shows the
progression of the matched predictions.
The right panel compares the N$^3$LL+N$^3$LO prediction with
fixed-order N$^3$LO.
The lower panels show ratios to the central N$^3$LL+N$^3$LO prediction.}
\label{fig:cumulantprogression}
\end{figure*}
\begin{table}[h]
\centering
\renewcommand{\arraystretch}{1.4}
\begin{tabular}{ccc}
\toprule
Order & $\sigma_{\text{N$^k$LO}} (30 \; \rm GeV)$ [pb] &  $\sigma_{\text{N$^k$LL}+\text{N$^k$LO}} (30 \; \rm GeV)$  [pb]  \\
\hline
0                    & $16.16^{+27\%}_{-19\%}$  & $ 15.35^{+26\%}_{-19\%}$\\
1                    & $25.86^{+14\%}_{-12\%}$  & $ 24.32^{+15\%}_{-13\%}$\\
2                 & $28.69^{+3.5\%}_{-5.9\%}$ & $28.98^{+5.9\%}_{-7.7\%}$ \\
3              & $29.24^{+0.7\%}_{-2.2\%}$  &  $30.11^{+3.2\%}_{-3.0\%}$ \\
\bottomrule
\end{tabular}
\caption{Cross section for Higgs production via gluon fusion
with a veto on the leading jet, $p_T^{\rm v} = 30\,\text{GeV}$.}
\label{tab:jetveto}
\end{table}
%

It is instructive to begin by recalling the previous state-of-the-art prediction at NNLL+N$^3$LO, presented in Ref.~\cite{Banfi:2015pju}. The latter also includes finite heavy-quark-mass effects and small-$R$ resummation at LL, both of which were found to be at the $1$-$2\%$ level~\cite{Banfi:2015pju,Banfi:2013eda} and are not included here.
Fig.~\ref{fig:N3LOcorrection} compares the NNLL+N$^3$LO and NNLL+NNLO predictions for the jet-veto cross section. 
At large $p_T^{\rm v}$, the inclusion of the N$^3$LO correction increases the cross section by $\sim\!3$--$4\%$ with respect to the NNLL+NNLO result, in line with the corresponding correction to the inclusive cross section~\cite{Anastasiou:2016cez}. The N$^3$LO correction, however, becomes anomalously small around $p_T^{\rm v}\sim30$ GeV, indicating a cancellation in the phenomenologically relevant region. 
In our multiplicative matching scheme~\cite{supplemental}, the difference between the two predictions involves leading power terms of ${\cal O}(\alpha_s^3\ln^m(M/p_T^{\rm v}))$ ($m\!=\!0,1$), and subleading power corrections. As we can see in Fig.~\ref{fig:N3LOcorrection}, this combination of terms vanishes around $p_T^{\rm v}\sim30$ GeV. 
It is natural to wonder if the cancellation can be lifted by the inclusion of N$^3$LL resummation, which precisely involves the ${\cal O}(\alpha_s^3\ln(M/p_T^{\rm v}))$ terms at this order.

Fig.~\ref{fig:cumulantprogression} summarizes the impact of the N$^3$LL resummation.
The left panel shows the matched progression at N$^m$LL+N$^m$LO, with $m=1,2,3$. 
As expected, the N$^3$LL resummation resolves the cancellation observed in Fig.~\ref{fig:N3LOcorrection} and restores a more uniform third-order correction (e.g., NNLL+NNLO vs.~N$^3$LL+N$^3$LO) of about $3$--$4\%$ for $p_T^{\rm v}\gtrsim20$ GeV.
The substantial overlap between the NNLL+NNLO and
N$^3$LL+N$^3$LO bands indicates perturbative
stability, and the figure shows a marked reduction of the uncertainty at the highest accuracy. At large $p_T^{\rm v}$, the separation between the different orders increasingly reflects the corresponding fixed-order inclusive normalizations.
The right panel of Fig.~\ref{fig:cumulantprogression} compares the N$^3$LL+N$^3$LO and N$^3$LO results. In the region $p_T^{\rm v}\simeq20$--$30$ GeV, resummation shifts the N$^3$LO prediction upward and produces a wider perturbative uncertainty band, reflecting the additional sensitivity to logarithmic higher-order corrections.

The noticeable N$^3$LL correction is consistent with the large three-loop rapidity anomalous dimension~\cite{companion}. This contribution constitutes genuinely new perturbative information and cannot be inferred from the moderate size of the NNLL correction or from the corresponding scale variations. The sizable N$^3$LL corrections are also reflected in the residual perturbative uncertainty, which remains at the $3\%$ level for veto scales of phenomenological interest.

Table~\ref{tab:jetveto} reports cross-section values at $p_T^{\rm v}=30$ GeV. The fixed-order cross section increases from $28.69~\mathrm{pb}$ at NNLO to $29.24~\mathrm{pb}$ at N$^3$LO, corresponding to a correction of about $2\%$, while the scale uncertainty decreases to the $2\%$ level.
The inclusion of N$^3$LL resummation raises the N$^3$LO result by a further $\sim\!3\%$, to $30.11^{+3.2\%}_{-3.0\%}~\mathrm{pb}$. This shift is not covered by the narrow N$^3$LO scale-variation band, showing that fixed-order scale variations alone do not capture the missing logarithmic corrections at phenomenologically relevant veto scales.
Relative to NNLL+NNLO, the N$^3$LL+N$^3$LO central value increases by about $4\%$, while the perturbative uncertainty decreases from approximately $6$--$8\%$ to about $3\%$.\footnote{An implementation of our calculation in the code of Refs.~\cite{Banfi:2012jm,Banfi:2015pju} is available at \url{https://github.com/pmonni/JetVHeto}}

\paragraph*{Conclusions.---}
We have presented the first prediction for the Higgs jet-vetoed cross section at N$^3$LL+N$^3$LO accuracy in QCD.
At $p_T^{\rm v}=30$ GeV, the N$^3$LL+N$^3$LO result is approximately $4\%$ larger than NNLL+NNLO and about $3\%$ larger than fixed-order N$^3$LO. The size of this correction could not have been inferred from the modest impact of NNLL resummation nor from a subset of the N$^3$LL ingredients, highlighting the importance of a complete calculation for precision phenomenology.
Our final prediction has a perturbative uncertainty of about $3\%$ for phenomenologically relevant veto scales. 
This precision matches that targeted for exclusive zero-jet cross section measurements at the high-luminosity LHC and is therefore directly relevant to Higgs-coupling determinations.

Future phenomenological applications should incorporate $t$- and $b$-quark mass effects~\cite{Banfi:2013eda,Banfi:2015pju,Caola:2018zye} and investigate the resummation of logarithms of the jet radius~\cite{Dasgupta:2014yra,Banfi:2015pju}. Although leading-logarithmic small-$R$ effects were found to be at the
percent level for $p_T^{\rm v}\simeq30~\mathrm{GeV}$~\cite{Banfi:2015pju}, the relatively
large subleading $\ln(R)$ contributions to the three-loop rapidity
anomalous dimension~\cite{companion} motivate investigating their
resummation. The reduction of $\gamma_\nu^{(3)}(R^2)$ at larger $R$~\cite{companion} is consistent with this interpretation.

\paragraph*{Acknowledgments.---}
We are grateful to Alex Huss for assistance with the running of the \texttt{NNLOJet} program. We also wish to thank Andrea Banfi, Thomas Becher, Gavin Salam, and Giulia Zanderighi for constructive comments on the article.
SA is
supported by the Ram\'on y Cajal program under grant
RYC2024-048348-I.
Part of the work of JRG has been supported by the Royal Society through Grant 
URF\textbackslash{}R1\textbackslash{}201500. 
The work of RS has been supported by the U.S. Department of Energy under Contract No. DE-SC0012704. 
This research was supported in part by grant NSF PHY-2309135 to the Kavli Institute for Theoretical Physics (KITP).
The work of PM is funded by the European Union (ERC, grant agreement
No. 101044599, JANUS).
Views and opinions expressed are however those of the authors only
and do not necessarily reflect those of the European Union or the
European Research Council Executive Agency. Neither the European
Union nor the granting authority can be held responsible for them.
We thank the Erwin-Schr\"odinger International Institute for Mathematics and Physics at the University of Vienna for partial support during the Programmes ``Quantum Field Theory at the Frontiers of the Strong Interactions", July 31 - September 1, 2023 and ``New Paradigms for Harnessing Quantum Field Theory at Colliders", July 27 - August 28, 2026.

\bibliographystyle{apsrev4-2}
\bibliography{refs}

\newpage

\onecolumngrid
\newpage
\appendix

\makeatletter
\renewcommand\@biblabel[1]{[#1S]}

\renewcommand{\theequation}{S.\arabic{equation}}
\setcounter{equation}{0}
\setcounter{figure}{0}
\renewcommand{\thefigure}{S\arabic{figure}}
\makeatother

\section*{Supplemental material}
Here we report additional technical details about the matching used to obtain the results presented.
%
%
The central predictions obtained in the Letter are derived according to the following multiplicative matching scheme. The resummation is switched off at large values of $p_T^{\rm v}$ using the procedure of Ref.~\cite{vanBeekveld:2025zjh}, which relies on a modification of the resummation scale as a function of $p_T^{\rm v}$. To present the matching scheme, we report the expressions of the resummed cross section, namely
\begin{align}
    \frac{d\sigma_{\text{N$^k$LL}}(p_T^{\rm v})}{d y} &= \hat{\sigma}_{\rm B} {\cal H}\left(M;M\right)\, e^{-\mathcal R\left(M,p_T^{\rm v}\right)}\,{\cal F}\left(M,p_T^{\rm v},R^2\right)\, \left({\cal C}\otimes f\right)_g(x_n,p_T^{\rm v}) \left({\cal C}\otimes f\right)_g(x_{\bar{n}},p_T^{\rm v})\,,
\end{align}
and of its fixed order counterpart, differential in the Higgs boson's rapidity $y$
\begin{equation}
   \frac{d\sigma_{\text{N$^m$LO}}(p_T^{\rm v})}{d y} = \frac{\sigma_{\text{N$^m$LO}}^{\rm tot.}}{dy} - \frac{d\sigma^{H+\rm jet}_{\text{N$^{m-1}$LO}}(p_T^j>p_T^{\rm v})}{dy}\,.
\end{equation}
We combine the two distributions as follows:
\begin{align}\label{eq:scheme-default}
    \frac{d\sigma_{\text{N$^k$LL+N$^m$LO}}(p_T^{\rm v})}{d y} &= e^{-\mathcal R\left(M,p_T^{\rm v}\right)}\,{\cal F}\left(M,p_T^{\rm v},R^2\right)\,\Bigg\{\sigma_{\rm B} {\cal H}\left(M;M\right)\, \left({\cal C}\otimes f\right)_g(x_n,p_T^{\rm v}) \left({\cal C}\otimes f\right)_g(x_{\bar{n}},p_T^{\rm v})\notag\\
    & + \Bigg[\frac{e^{\mathcal R\left(M,p_T^{\rm v}\right)}}{{\cal F}\left(M,p_T^{\rm v},R^2\right)}\left(\frac{d\sigma_{\text{N$^m$LO}}(p_T^{\rm v})}{d y}-\frac{d\sigma_{\text{N$^k$LL}}(p_T^{\rm v})}{d y}\right)\Bigg]_{\text{Expanded through N$^m$LO}}\Bigg\}\,,
\end{align}
where the second term in the curly braces is to be expanded in powers of $\alpha_s$ at the same N$^m$LO order.
At the nominal order of N$^3$LL+N$^3$LO, this scheme allows the extraction of the missing N$^3$LO leading-power non-logarithmic terms, of formal N$^4$LL order, from the fixed-order calculation.

\begin{figure*}[ht]
\includegraphics[width = 0.45\textwidth]{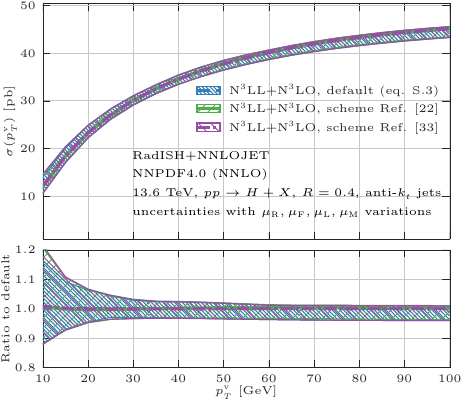}
\includegraphics[width = 0.45\textwidth]{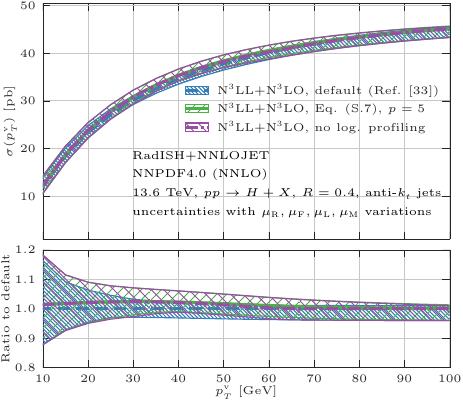}
\caption{Jet-vetoed cross section at $\sqrt{s}=13.6~\mathrm{TeV}$ as a function of the veto scale $p_T^{\rm v}$, for anti-$k_t$ jets with radius parameter $R=0.4$ at N$^3$LL+N$^3$LO accuracy using different matching schemes (left panel) and using different prescriptions for turning off the resummation at large $p_T^{\rm v}$ (right panel).}
\label{fig:matchlogunc}
\end{figure*}

To assess the uncertainty associated with the choice of the matching scheme, we also perform the matching using the alternative schemes of Refs.~\cite{Banfi:2015pju,vanBeekveld:2025zjh}. These are defined similarly to Eq.~\eqref{eq:scheme-default}, namely
\begin{align}\label{eq:scheme-master}
    \frac{d\sigma_{\text{N$^k$LL+N$^m$LO}}(p_T^{\rm v})}{d y} &= \frac{1}{\cal N}\frac{d\sigma_{\text{N$^k$LL}}(p_T^{\rm v})}{d y}\,\Bigg\{{\cal N} + \Bigg[\frac{\cal N}{\frac{d\sigma_{\text{N$^k$LL}}(p_T^{\rm v})}{d y}}\left(\frac{d\sigma_{\text{N$^m$LO}}(p_T^{\rm v})}{d y}-\frac{d\sigma_{\text{N$^k$LL}}(p_T^{\rm v})}{d y}\right)\Bigg]_{\text{Expanded through N$^m$LO}}\Bigg\}\,,
\end{align}
where
\begin{align}
    \mathcal{N} &=  \hat{\sigma}_{\rm B} {\cal H}\left(M;M\right)\,\left({\cal C}\otimes f\right)_g(x_n,p_T^{\rm v}) \left({\cal C}\otimes f\right)_g(x_{\bar{n}},p_T^{\rm v})\frac{f_g(x_n,M)\,f_g(x_{\bar{n}},M)}{f_g(x_n,p_T^{\rm v})\,f_g(x_{\bar{n}},p_T^{\rm v}) }\,,\\
    {\mathcal {N}} &= \left. \frac{d\sigma_{\text{N$^k$LL}}(p_T^{\rm v})}{d y}\right|_{L\to 0}\,,
\end{align}
for the schemes of Refs.~\cite{Banfi:2015pju} and~\cite{vanBeekveld:2025zjh}, respectively. Our default procedure to switch off the resummed logarithms $L = \ln \frac{\mu_L}{p_T^{\rm v}}$ (with $\mu_L$ being the resummation scale) at large $p_T^{\rm v}$ is taken from Ref.~\cite{vanBeekveld:2025zjh}. Finally, we also investigated the uncertainty associated with the way the resummation is turned off at large $p_T^{\rm v}$ by considering, alongside the default profiled-resummation-scale procedure from Ref.~\cite{vanBeekveld:2025zjh}, the modified logarithms (cf., e.g., Ref.~\cite{Banfi:2015pju})
\begin{equation}
 L\to \tilde L = \frac{1}{p} \ln \left[ \left(\frac{\mu_L}{p_T^{\rm v}}\right)^p + 1\right]\,,
\end{equation}
(where now $\mu_L$ is not profiled) and the use of unmodified logarithms 
\begin{equation}
 L\to  L\times\Theta(\mu_L- p_T^{\rm v})\,.
\end{equation}
Note that there is no matching scale for the latter two options.
We display results at N$^3$LL+N$^3$LO accuracy with different matching schemes (using our default choice for switching off the logarithms) and with different ways of turning off the logarithms (for our default matching scheme) in the left and right panel of Fig.~\ref{fig:matchlogunc}. In all cases, the results are consistent within the quoted uncertainties of our default prediction.

\begin{figure*}[ht]
\includegraphics[width = 0.45\textwidth]{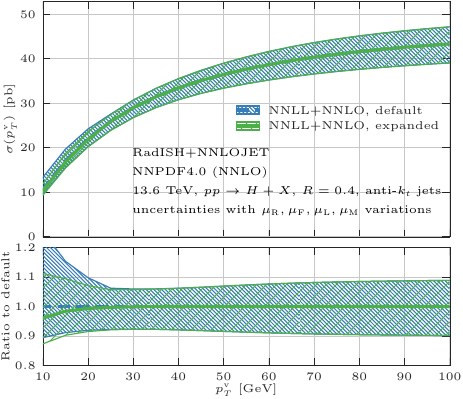}
\includegraphics[width = 0.45\textwidth]{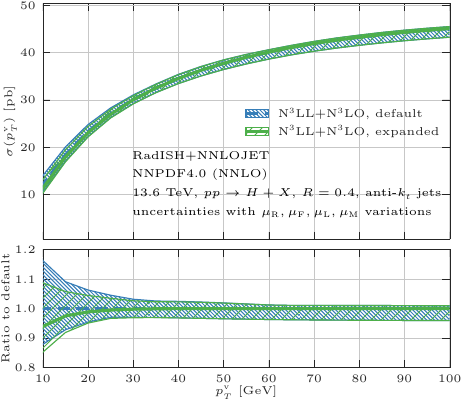}
\caption{Comparison of the jet-vetoed cross section at $\sqrt{s}=13.6~\mathrm{TeV}$ at NNLL+NNLO (left panel) and N$^3$LL+N$^3$LO (right panel) accuracy using two different treatments of the $R$-dependent corrections in Eq.~\eqref{eq:FF}.}
\label{fig:exp-vs-noexp}
\end{figure*}

To conclude, in Fig.~\ref{fig:exp-vs-noexp} we also show the comparison of the default predictions at NNLL+NNLO and N$^3$LL+N$^3$LO to the ones obtained by expanding the exponentiated jet-algorithm-dependent term of the resummation, Eq.~\eqref{eq:FF}, at the quoted logarithmic accuracy and neglecting subleading logarithmic effects. The latter prescription differs from Eq.~\eqref{eq:FF} by subleading logarithmic terms, and it is a measure of further subleading corrections. We observe that the two prescriptions agree within the quoted uncertainty band. Together with the results shown in Figs.~\ref{fig:matchlogunc}, this indicates the robustness of our uncertainty estimate.

%

\end{document}